\documentstyle[epsfig,longtable]{aipproc}
\newcommand{\degs}{\mbox{\(^\circ \)}}
\newcommand{\gr}{$\gamma$-ray \,}
\newcommand{\grs}{$\gamma$-rays \,}

\begin{document}
\title{HEGRA Observations of Galactic Sources}
\author{Heinz V\"olk$^*$, HEGRA Collaboration}
\address{$^*$Max-Planck-Institut f\"ur Kernphysik\\
P.O. Box 103980, 69029 Heidelberg, Germany}
\maketitle
\begin{abstract}
In this talk I will first give a summary of the observations of expected
Galactic TeV \gr sources with the HEGRA CT-Sytem since the Kruger Park
Workshop in 1997. Then I will go into some detail regarding the
observations of Supernova Remnants (SNRs), especially those of Tycho's SNR
and of Cas~A. The emphasis will not be on all aspects of these published
data. I will rather review the selection of these observational targets,
and discuss some of the physical implications of the results.
\end{abstract}
\section*{Summary of observations}
The stereoscopic system of imaging atmospheric Cherenkov telescopes
(IACTs) of HEGRA has been running since late 1996 with four telescopes.
After a fire in the array which also damaged one of these telescopes late
in the year 1997, the final configuration of five equal
telescopes
with
identical cameras has become operational in August 1998. Apart from the
IACT system, HEGRA successfully operates a stand-alone telescope, called
CT1; it is also doing obervations during moon periods. However in this
review, I will be concerned with the stereoscopic system alone.

Since 1997 a number of Galactic source candidates has been observed with
the IACT system. The Galactic coordinates of the objects discussed in this
paper are indicated in Figure 1. The objects analyzed are given in Table
1.\\
1. The observations of the Crab Nebula were done both at normal (ZA
$\leq 30$\degs) and at high zenith angles (ZA $\sim 60$\degs). They led to
a (combined) energy spectrum up to 20 TeV (Konopelko et al. 1999). It is
within the errors compatible with an extension of the power law spectrum
inferred from measurements in the TeV energy range (Konopelko et al.,
these Proceedings). Thus no possible hard hadronic emission component has
been identified up to these energies.\\
2. A search for a periodic signal from the Crab and Geminga   
Pulsars was also performed (Aharonian et al. 1999a). No evidence for 
pulsed emission was found. Even though we had expected Geminga to be
a major contributor to the distribution of very energetic CR electrons
in the neighborhood the Solar System, it showed up as a TeV-quiet
object.

\begin{figure}[t] 
  \centerline{\epsfig{file=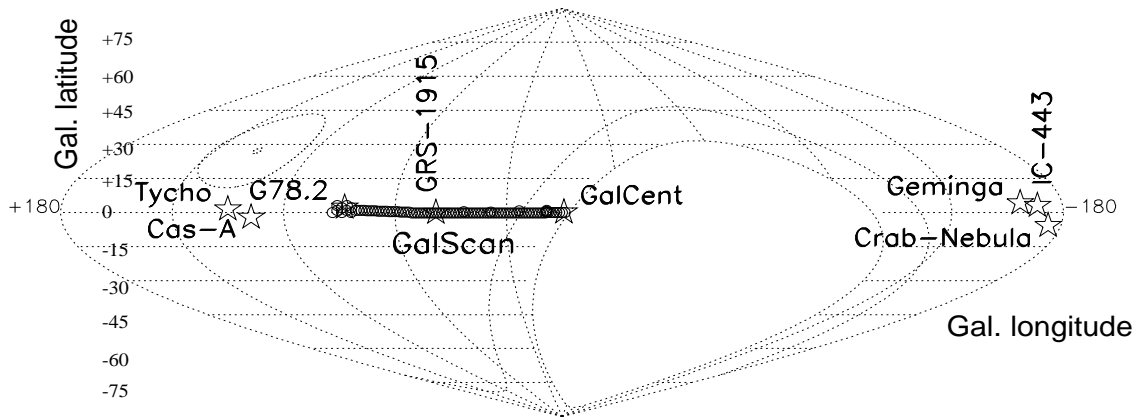,height=60 mm,width=154 mm}}
  \vspace{10pt}
  \caption{Part of the sky in Galactic coordinates.  Shown
    are the positions of the objects discussed in this paper. The white part
    is not visible from La Palma; the band surrounding that region can only be   
    observed under restricted conditions.}   
  \label{fig1}
\end{figure}

\begin{table}

\caption{Galactic objects observed 1997-1999 with the HEGRA IACT-System.
The numbers given are the observation hours in the respective years
for the analysis of the Galactic Cosmic-Ray (CR) proton spectrum, CR 
background (bgr) events were used.}
\begin{tabular}{ldddd}
Source & [1997] & [1998] & [1999] & References \\
\tableline
 Crab       &  92  & 138 &  31 & Konopelko et al. 1999, \\
            &      &     &     &  Aharonian et al. 1999a \\
 Geminga    &  --  &  23 &  -- & Aharonian et al. 1999a   \\
 GRS 1915   &  50  &  12 &  11 & Kettler 1999            \\
 Cas~A      & 102  &  85 &  -- & P\"uhlhofer et al. 1999a   \\
 Tycho      &  23  &  35 &  -- & P\"uhlhofer et al. 1999a  \\
 Gal. Plane & 111  &  66 &  -- & P\"uhlhofer et al. 1999b  \\
 Diff. VHE  &  --  &  -- &  53 & Lampeitl et al. 1999  \\
 CR Protons & bgr  & bgr & bgr & Aharonian et al. 1999b \\
\end{tabular}
\end{table}

\noindent 3. The observations of the Galactic Microquasar GRS 1915 have
been
analysed (Kettler 1999). No signals have been found during these
observation periods. This is not too surprising since, unfortunately, the
source had also been low in other wavelength ranges in those times.\\
4. The two SNRs Tycho and Cas~A have been observed extensively with
the
stereoscopic system. This is especially true for Cas~A, where a deep
observation of 128 hr duration has been included in the present analysis
(P\"uhlhofer et al. 1999a). The two objects will be discussed in some
detail in section 2.\\
5. An extensive Galactic Plane scan ($\geq 2$ hrs of observation
time for each point,
plus
some re-observations) in the TeV band covered the
 Galactic longitude region from the Galactic Center ($l,-1.5 \degs$) to
the
Cygnus region ($l,83.5 \degs$), see Figure 2 .
\begin{figure}[t] 
\centerline{\epsfig{file=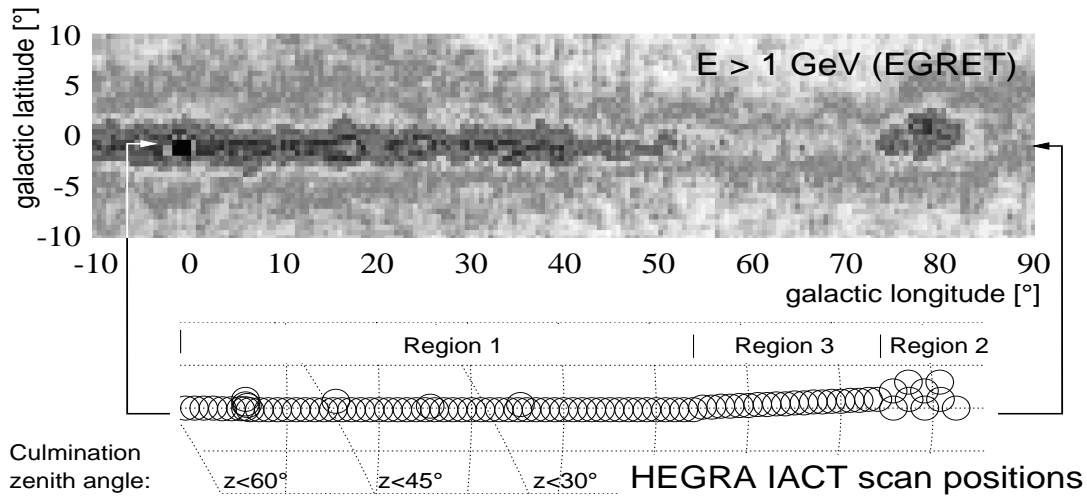,height=74 mm,width=150
mm}}  
\vspace{10pt}
\caption{The HEGRA IACT scan positions in the Galactic Plane.
For comparison also the corresponding part of the EGRET sky map for \gr 
energies $E>1{\rm GeV}$ is shown. Regions 1 and 2 were observed in 1997,
region 3 in 1998.}
\label{fig2}
\end{figure}


%
%

\noindent Sources with a flux above 1/4 Crab units should have been
detected, as
indicated by Table 2 below. A first analysis reveals no hints for such
strong TeV point sources (P\"uhlhofer et al. 1999b). \\
6. In a similar vein, a program regarding the search for diffuse VHE
\gr
emission from the Galactic Plane was started. The present analysis is
largely of a technical nature (Lampeitl et al. 1999). The observations
will be continued during this summer of 1999.\\
7. Finally, the imaging Cherenkov technique was applied for the first time
to
the determination of the flux and the TeV energy spectrum of the
charged CR protons. For this work background events for the Mkn 501
observations from 1997 were used. Calibration is exclusively by Monte
Carlo simulations that include a detailed detector simulation. For
physical reasons, the proton detection rate strongly exceeds that for
heavier CR nuclei near threshold, around 1.5 TeV. The stereoscopic
detection of the air showers permits the effective suppression of air
showers induced by heavier particles already at the trigger level, and in
addition by software analysis cuts. The results are in good agreement with
the recent results of satellite and balloone-borne experiments and reach
similar accuracy (Aharonian et al. 1999b). Without any knowledge of
the CR
composition, the proton spectrum can only be determined with precision
near threshold. However, it should be possible to obtain in addition an
approximate CR composition, using further specific image cuts
(Plyasheshnikov et al. 1998). This will allow an extension of the
dynamical range of the spectrum into an energy region that is very costly
to cover by direct detection CR experiments. I think it would be important
if in addition the large Zenith angle technique could be applied to this
problem.\\
Of course, more than these objects have been observed in the Galaxy.
However the data have not been analyzed yet, and are therefore not a
subject of this summary.

Let me conclude this section with a general consideration. \\
As mentioned above, the HEGRA Galactic Plane scan has not yet led to the
detection of new sources. In his excellent introductory review, Trevor
Weekes (these Proceedings) described this result as "depressing". \\
We were also disappointed. On the other hand, the
result is
perhaps not too surprising, given the low sensitivity level with which
this survey had to be done. The result should also prompt a new discussion
about the aims and possibilities of ground-based \gr astronomy with
imaging
telescopes. Space is scarce in these proceedings. So, I will summarize my
arguments only briefly in four points, and hope that they open a
broader debate: (i) we should of course continue such surveys; any
field of astronomy must do this (ii) however it is not too probable that
we will find new sources that have not been seen as unusual objects in
another wavelength range already, considering the enormous investments in
ever more powerful instruments in the radio, infrared, optical, and X-ray
domains that have been made over the last two decades (iii) thus, our main
activity should perhaps be to look at sources also known in other
wavelength-ranges; only then we can hope to obtain a physical
understanding of the \gr results (iv) given the much higher physical
complexity of the acceleration and transport processes for the nonthermal
component than for the thermal component, the potential for discovery is
one for strong nonthermal activity in known objects, and it is as
important as the potential for discovery of previously unknown
objects in the more conventional "thermal" astronomy.\\
I do not believe that the \gr bursts provide a counter argument to this
point of view: they are explosive events in previously inconspicuous
objects, and could not have been found in a survey with a narrow-FoV
instrument like IACTs; an all-sky capability was needed to discover them,
and they were difficult to understand for decades before they were
detected also in other wavelength ranges. Also Geminga, originally an
enigmatic Cos~B source, is not really a counterexample, because Geminga
could only be physically identified after many years, when ROSAT
discovered that it was a long-period Pulsar and determined its period,
which was subsequently confirmed by EGRET.

\section*{Supernova Remnants}
\subsection*{Observations}
Earlier observations of the SNRs G87.2+2.1 ($\gamma$~Cygni) and IC~443 in
1996/97 gave consistent upper limits between Whipple (Buckley et al.
1998) and the HEGRA CT-System (He\ss ~1998) at effective threshold
energies, for the Zenith angles involved, of $E_{\gamma} > 300$~GeV and
$E_{\gamma} > 800$~GeV, respectively. They were slightly above
theoretical predictions regarding the $\pi^0$-decay \gr emission for a
uniform ISM but well within astronomical uncertainties (V\"olk 1997).
Both objects had originally been assumed to interact with interstellar
clouds. Under ideal assumptions such an interaction could have increased
the $\pi^0$-decay \gr luminosity significantly.\\
These two SNRs are presumably the result of core collapse Supernovae, due
to massive ($M>8\, {\rm M}_{\odot}$) progenitor stars. If they have masses
exceeding roughly $15~M_{\odot}$, these stars have stellar winds which
significantly modify the circumstellar environment. For such "Wind-SNe"  
the time history of the $\pi^0$-decay \gr emission is much more complex:
except within the wind zone, it is much lower than for a uniform ISM
of the same density (Berezhko \& V\"olk 1995, Berezhko \& V\"olk 1997).

The recent HEGRA observations concern deep observations of Tycho's SNR,
believed to be a SN Ia in a uniform ISM with strong X-ray lines and no or
only a very weak nonthermal X-ray continuum, and of Cas~A, assumed to be a
SN Ib resulting from the core collapse of a very massive Wolf-Rayet star,
with a strong nonthermal X-ray continuum - an archetypical Wind-SN  
(P\"uhlhofer et al. 1999a). Both SNRs are very young in
an evolutionary
sense, presumably still in the sweep-up phase, even though this might only
be marginally true for Tycho's SNR.

The long observation times which we have reserved for these objects imply
a change in our ideology: the emphasis is no more on SNR shocks that
presumably interact with interstellar clouds, but on very young objects,
either in a supposedly uniform ISM or in a strongly modified precursor
wind structure.

\subsection*{Data analysis for Cas~A and Tycho}
The following data analysis for Tycho and Cas~A has been done by G.
P\"uhlhofer. For the angular resolution of the HEGRA CT-system of $0.05~
{\rm to}~
0.1 \degs$ Tycho is a \gr point source, and Cas~A is marginally extended.
Due to the available Zenith angles the instrument threshold is at 1 TeV or
slightly above (see Table 2). The complete sensitivities of the IACT
system are described in the article by M. Panter (these Proceedings).
The observation times and significances are given in Table 2.
\begin{figure}[t] 
\centerline{\epsfig{file=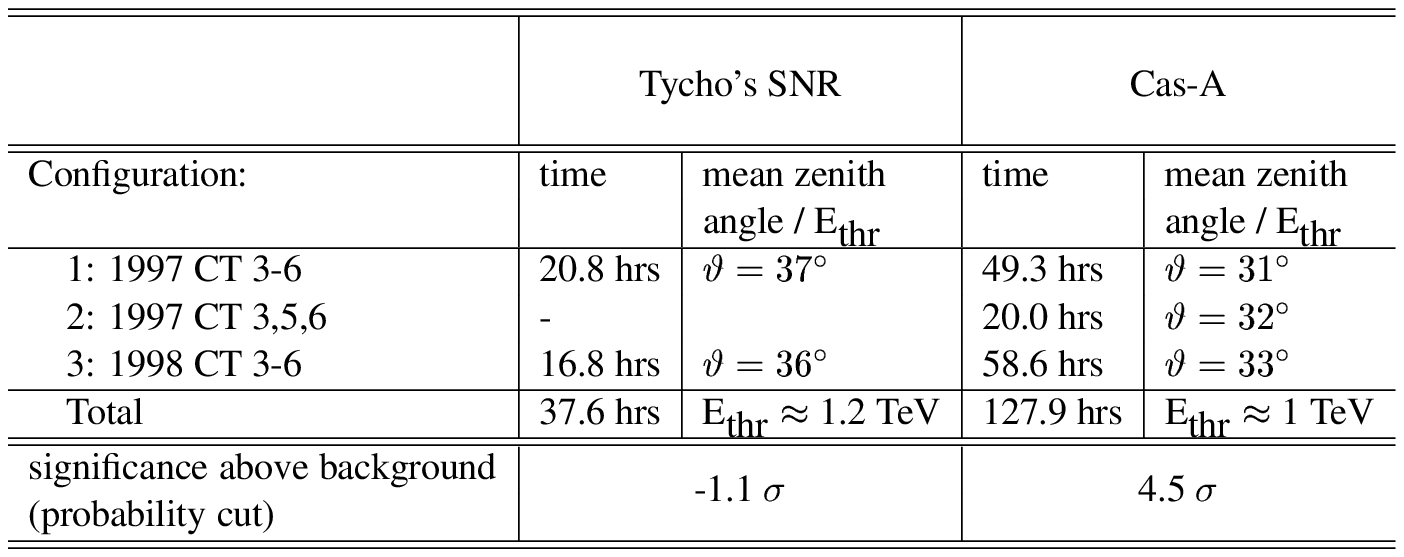,height=55mm,width=140mm}}
\vspace{10pt}
\caption{Observation times and significances for
Tycho's SNR
and Cas~A;
the 37.6 hrs for Tycho constitute only part of the total observation time
available. The stereoscopic configuration of the four sytem telescopes CT
3-6, used for these observations, was different for the second 1997 period
due to the fire on La Palma that hit CT 4. For the analysis of Cas~A the
source was assumed to be slighly extended.}
\label{Table2}
\end{figure}

With $\sim 38$ hrs of observation no signal has yet been found from Tycho,
whereas the full data sample of $\sim 128$ hrs for Cas~A shows evidence
for a signal above 1 TeV.
For Cas~A the event statistics as a function of $({\rm distance})^2$ is
shown in the left panel of Figure 3
, both for a point source
assumption (I), and for a slightly extended source (II). The position of
the \gr source on the sky is given in the right panel, and is consistent
with the radio astronomical position that corresponds to the center of the
picture.
\begin{figure}[t] 
\centerline{\epsfig{file=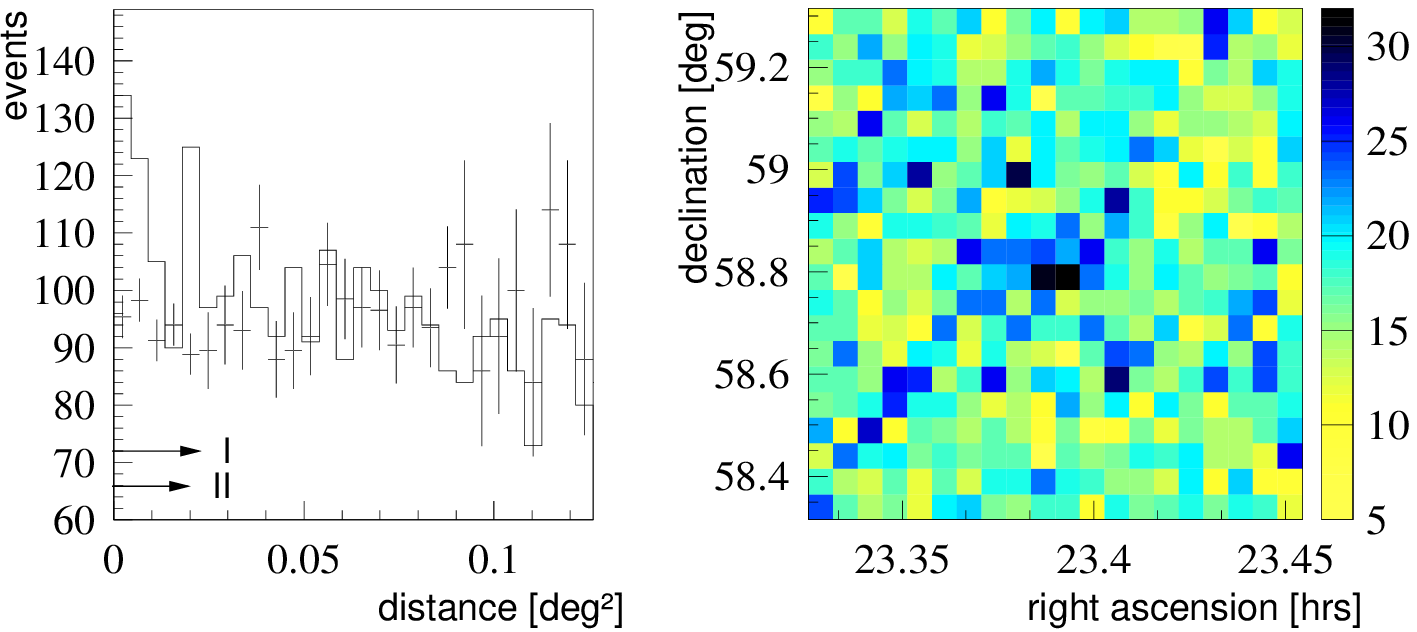,height=63mm,width=143mm}}
\vspace{10pt}
\caption{Event statistics and position determination of
the \gr source Cas~A. See text for details.}
\label{fig3}
\end{figure}

Figure 4 
shows a model calculation for the energy spectrum of
Cas~A (Atoyan et al. 1999b). The full and the dashed lines
correspond to
the expected inverse Compton (IC) emission, as derived phenomenologically
from the observed synchrotron spectrum from the radio to the hard X-ray
region (Atoyan et al. 1999a), for the two mean magnetic field
strengths
$B_1=1 {\rm mG and}=1.6 {\rm mG}$, respectively. The heavily dotted curve
corresponds to a $\pi^0$-decay spectrum, produced by an assumed power
law
spectrum of protons accelerated in the source, with a total energy content
of $W_{p}=2 \times 10^{49}{\rm ergs}\simeq 1/5 W_{p}(t=\infty)$, where
$W_{p}(t=\infty)=10^{50} {\rm erg}$ corresponds to an assumed
time-asymptotic nonthermal fraction of 10 percent of the total
hydrodynamic energy of $10^{51} {\rm erg}$, generally assumed to be
released in Cas~A.
The proton spectral index assumed is 2.15, with a rather high cutoff at
200 TeV for this Wind-SN already at very early times (V\"olk \& Biermann
1988). The
mean thermal gas density seen by the relativistic protons is taken as $15~
{\rm cm}^{-3}$. The energetic electrons responsible for the IC emission
are assumed to come from three different regions of the SNR interior: the
bright compact radio components with magnetic field strength $B_1=1~{\rm
mG}$, where electrons are accelerated locally, an extended "plateau" of
shocked circumstellar gas with $B_2=B_1/4$ due to global acceleration at
the forward SNR shock, and a low-field part of this "plateau" with
$B_3=0.1~{\rm mG}$. The electron spectral index is assumed to be uniformly
2.15, as for the protons. The electron cutoff energy, however, is only 17
TeV, corresponding to the steep drop-off of the observed hard X-ray
spectrum with increasing energy.
%
\begin{figure}[b!] 
\centerline{\epsfig{file=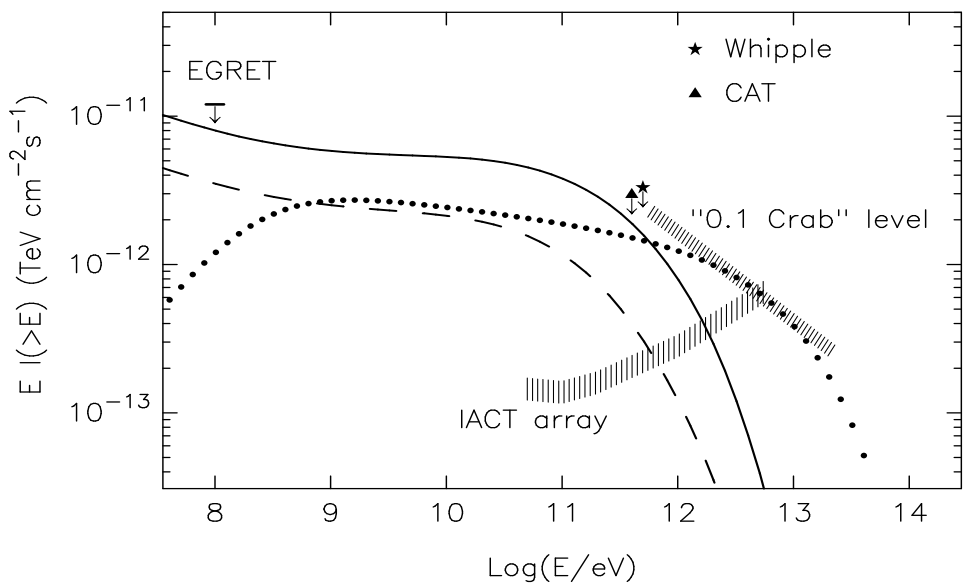,height=3.0in,width=5.0in}} 
\vspace{10pt}
\caption{Energy flux spectrum of Cas~A in \grs. The
full and the dotted curve correspond to the inferred IC emission, for a
magnetic field strength $B_1$ in the bright compact radio knots, given by
$B_1=1 {\rm mG~ and}=1.6 {\rm mG}$, respectively. The heavily dotted curve
is an assumed $\pi^0$-decay spectrum thought to be appropriate for the
present evolutionary state of the remnant (see text). Indicated are also
the upper limits reported by the Whipple (Lessard et al. 1999) and
CAT
collaborations (Goret et al. 1999). The slantedly hatched curve
denotes
1/10 of the Crab flux. The vertically hatched curve corresponds to the
sensitivity of the future arrays Cangaroo III, H.E.S.S. and VERITAS.}
\label{fig4} 
\end{figure}

Clearly, the magnitude of the \gr flux at about 1 TeV, if ultimately
detected with a significance exceeding $5 \sigma$, could be equally due to
electronic IC or hadronic $\pi^0$-decay emission. However the spectra
would
be very different for the two cases: an IC spectrum should fall off
strongly with energy, in contrast to a $\pi^0$-decay spectrum. 
Therefore I
believe that every effort should be made to obtain a TeV-spectrum of
Cas~A.

{\bf Acknowledgements} I am grateful to Gerd P\"uhlhofer for providing
several of the Figures in this paper.


\begin{references}
%
\bibitem{Konopelkoetal99}Konopelko, A.K., P\"uhlhofer, G., et al., {\it
Proc. 26th ICRC, Salt Lake City} {\bf 3}, 444 (1999).
%
\bibitem{Aharonianetal99a}Aharonian, F.A., Akhperjanian, A.G., Barrio,
J.A., et al., {\it A\&A} {\bf 346}, 913 (1999a).
%
\bibitem{Kettler99}Kettler, J., private communication, (1999).
%
\bibitem{Puhlhofer99a}P\"uhlhofer, G., V\"olk, H.J., Wiedner,
C.-A., et al.,
{\it Proc. 26th ICRC, Salt Lake City} {\bf 3}, 492 (1999).
%
\bibitem{Puhlhofer99b}P\"uhlhofer, G., Bernl\"{o}hr, K., Daum, A.,
et
al., {\it Proc. 26th ICRC, Salt Lake City} {\bf 4}, 77 (1999).
%
\bibitem{LampeitlK99}Lampeitl, H., Konopelko, A.K., et al.,
{\it Proc. 26th ICRC, Salt Lake City} {\bf 4}, 81 (1999).
%
\bibitem{Aharonianetal99b}Aharonian, F.A., Akhperjanian, A.G., Barrio,
J.A., et al., {\it A\&A} {\bf 59}, 092003-1 (1999b).
%
\bibitem{Plyasheshnikovetal99b}Plyasheshnikov, A.V., Konopelko,
A.K., Aharonian, F.A., et al., {\it J. Phys. G} {\bf 24}, 653
(1998).
%
\bibitem{BuckleyAC98}Buckley, J.H., Akerlof,  C.W., Carter-Lewis,
D.A., et al., {\it A\&A} {\bf 329}, 639 (1999).
%
\bibitem{Hess98}He\ss, M., {\it PhD Thesis Univ. Heidelberg} (1998).
%
\bibitem{Volk97}V\"olk, H.J., {\it Proc. "Towards a Major
Atmospheric
Cherenkov Detector V", Kruger Park, S.A.}, 87 (1997).
%
\bibitem{BerezhkoV95}Berezhko, E.G., V\"olk, H.J.,
{\it Proc. 24th ICRC, Rome} {\bf 3}, 380 (1995).  
%
\bibitem{BerezhkoV97}Berezhko, E.G., V\"olk, H.J.,         
{\it Astroparticle Phys.} {\bf 7}, 183 (1997).          
%
\bibitem{AtoyanAT99a}Atoyan, A.M., Aharonian, F.A., Tuffs, R.J., et al.,
{\it A\&A} in press (1999a).
%
\bibitem{AtoyanAT99b}Atoyan, A.M., Aharonian, F.A., Tuffs, R.J., et al.,
{\it A\&A} submitted to {\it A\&A} (1999b).
%
\bibitem{VolkB88}V\"olk, H.J., Biermann, P.L., {\it ApJ} {\bf 333},
L65 (1988).
%
\bibitem{LessardBB99}Lessard, R.W., Bond, I.H., Boyle, P.J., et
al., {\it Proc. 26th ICRC, Salt Lake City} {\bf 3}, 488 (1999).
%
\bibitem{GoretGN99}Goret, P., Guiffes, C., Nuss, E., et
al., {\it Proc. 26th ICRC, Salt Lake City} {\bf 3}, 496 (1999).
\end{references}
\end{document}